\RequirePackage{lineno}
\documentclass[conference]{IEEEtran}
\usepackage{threeparttable}
\usepackage{lipsum}
\usepackage[dvips]{color}
\usepackage{graphicx}
\usepackage{lipsum}
\usepackage{algorithm,algorithmic}
\usepackage{epsfig,amsmath,amsfonts,cite}
\usepackage{booktabs}
\newcommand{\thickhline}{\noalign{\hrule height 1.0pt}}
\usepackage{multirow}%,multicol

\begin{document}
%\title{Uncertainty Quantification for the Periodic Steady States of Forced and Autonomous Circuits via Stochastic Testing and Shooting Newton}
%\title{Generating Positive Approximation for Circuit Performance Density Using Rational Function Fitting and Semi-Definite Programming}
\title{Uncertainty Quantification for Integrated Circuits: Stochastic Spectral Methods}
%\title{Uncertainty Quantification of Periodic Steady States for Analog, RF and Power Electronic Circuits}


\IEEEspecialpapernotice{(Invited Special Session Paper)}

\author{\IEEEauthorblockN{Zheng Zhang}
\IEEEauthorblockA{Research Lab of Electronics\\
Massachusetts Institute of Technology\\ 
Cambrige, MA 02139, USA\\
E-mail: z\_zhang@mit.edu}
\and
\IEEEauthorblockN{Ibrahim (Abe) M. Elfadel}
\IEEEauthorblockA{Microsystem Engineering\\ Masdar Inst. of Science \& Technology\\
Abu Dhabi, United Arab Emirates\\
E-mail: ielfadel@masdar.ac.ae}
\and
\IEEEauthorblockN{Luca Daniel}
\IEEEauthorblockA{Research Lab of Electronics\\ Massachusetts Institute of Technology\\
Cambrige, MA 02139, USA\\
E-mail: luca@mit.edu}

%\thanks{This work was supported by the MI-MIT Collaborative Program (Reference No.196F/002/707/102f/70/9374). Elfadel's work was also supported by SRC under the MEES I, MEES II, and ACE$^{4}$S programs, and by ATIC under the TwinLab program. Z. Zhang would like to thank Dr. Tarek El-Moselhy for his helpful discussions during the work of~\cite{zzhang:tcad2013,zzhang:tcas2_2013}}
}


%Telephone: (800) 555--1212, Fax: (888) 555--1212}


\maketitle

\begin{abstract}
%\boldmath
Due to significant manufacturing process variations, the performance of integrated circuits (ICs) has become increasingly uncertain. Such uncertainties must be carefully quantified with efficient stochastic circuit simulators. This paper discusses the recent advances of stochastic spectral circuit simulators based on generalized polynomial chaos (gPC). Such techniques can handle both Gaussian and non-Gaussian random parameters, showing remarkable speedup over Monte Carlo for circuits with a small or medium number of parameters. We focus on the recently developed stochastic testing and the application of conventional stochastic Galerkin and stochastic collocation schemes to nonlinear circuit problems. The uncertainty quantification algorithms for static, transient and periodic steady-state simulations are presented along with some practical simulation results. Some open problems in this field are discussed.
\end{abstract}


% \fi
\IEEEpeerreviewmaketitle

\section{Introduction}
Manufacturing process variations have led to significant performance uncertainties in submicron and nano-scale IC design~\cite{variation2008,VLSI2008}. Many results have been reported on variation-aware modeling for semiconductor devices~\cite{cicc2011,Boning00modelsof,Dopant}, interconnects~\cite{Tarek_DAC:08,Tarek_DAC:10,Tarek_ISQED:11,zzhang_iccad:2011,Shen2010,Wenjian:2009,Gong:2012,Hengliang:2007}, and for analog/RF and digital ICs~\cite{xli2010,TCAD2006}. However, few have focused on the uncertainty quantification aspect that analyzes the uncertainty propagation from the device level to the circuit level through SPICE simulation.

Monte Carlo (MC)~\cite{MCintro} has been the mainstream uncertainty quantification technique in commercial circuit simulators for decades~\cite{kundertbook:1995,Tuinenga:1995,Cadence,HSPICE}. Recently, Singhee {\it et al.} improved MC-based simulation and applied it to the yield analysis of analog/RF and digital ICs~\cite{SingheeR09,SingheeR10}. Despite its wide application, MC has a slow convergence rate proportional to $\frac{1}{\sqrt{N_s}}$ (where $N_s$ is the number of samples used in MC). Very often, one must run a huge number of SPICE simulations to achieve acceptable accuracy at a prohibitively high computational cost.

Stochastic spectral methods~\cite{sgscCompare,sfem,book:Dxiu,UQ:book,gPC2002,gPC2003,xiu2009} have emerged as a promising solution to uncertainty quantification problems, showing significant speedup over MC (especially when the parameter dimensionality is small or medium). Such methods represent the parameter-dependent solutions by some properly constructed basis functions, such as polynomial chaos (PC, also called Hermite polynomial chaos)~\cite{PC1938} or generalized polynomial chaos (gPC)~\cite{gPC2002}. Mainstream stochastic spectral solvers include the stochastic Galerkin (also called stochastic finite element method~\cite{sfem}) and stochastic collocation~\cite{col:2005,Ivo:2007,Nobile:2008,Nobile:2008_2} methods. Stochastic Galerkin is an intrusive (or ``non-sampling based") solver, since it directly computes the PC/gPC coefficients by solving a coupled equation resulting from Galerkin testing. Stochastic collocation is a non-intrusive (or ``sampling based") method: it solves the deterministic equations at a set of sample points, followed by a numerical scheme to reconstruct the PC/gPC coefficients. 

There is an increasing interest in applying stochastic spectral methods to circuit simulation. Most works use PC-based stochastic collocation or stochastic Galerkin methods to simulate on-chip and off-chip interconnects with Gaussian parameters~\cite{%Tarek_DAC:08,Tarek_DAC:10,Tarek_ISQED:11,
Stievano:2012,Stievano:2011_1,Fan:2007,Wang:2004,yzou:2007}. Limited results have been reported on nonlinear circuit analysis. The PC-based stochastic circuit simulator proposed by Strunz~\cite{Strunz:2008} requires constructing the stochastic device models {\it a-priori}, thus it cannot be easily integrated with industrial semiconductor device models. In~\cite{Tao:2007}, stochastic collocation was combined with harmonic balance to simulate nonlinear RF circuits under Gaussian variations.
\begin{table*}[t]
	\centering 
	\caption{Univariate gPC polynomial basis of some typical random parameters~\cite{xiu2009}.}	
	\label{tab:gPC}
	\begin{threeparttable}
	\begin{tabular}{|c||c|c|c|}
	\hline%\hline		
Distribution of $\xi_k$ & ${\rm PDF}$ of $\xi_k$ [$\rho_k(\xi_k)$]\footnotemark[1]& univariate gPC basis $\phi^k _{\nu} \left( {\xi _k } \right)$ & Support $\Omega _k$\\
\thickhline	
	Gaussian  & $\frac{1}{{\sqrt {2\pi } }}\exp \left( {\frac{{ - \xi _k^2 }}{2}} \right)$ &Hermite-chaos polynomial&  $(-\infty, +\infty)$ \\ \hline
	Gamma  & $\frac{{\xi _k ^{\gamma  - 1} \exp \left( { - \xi _k } \right)}}{{\Gamma \left( \gamma  \right)}},\;\gamma  > 0$  &Laguerre-chaos polynomial&  $[0, +\infty)$ \\ \hline
	Beta  & $\frac{{ {\xi_k}^{\alpha  - 1} \left( {1 - \xi_k} \right)^{\beta  - 1} }}{{{\rm B}\left( {\alpha ,\beta } \right)}},\;\;\alpha ,\beta  > 0 $ &Jacobi-chaos polynomial&  $[0, 1]$ \\ \hline
	Uniform  & $\frac{1}{2}$  &Legendre-chaos polynomial&  $[-1, 1]$ \\	
\hline%\hline
	\end{tabular} 
	\begin{tablenotes}
       \item[1] $\Gamma \left( \gamma  \right) = \int\limits_0^\infty  {t^{\gamma  - 1} \exp \left( { - t} \right)dt}$ and ${\rm B}\left( {\alpha ,\beta } \right) = \int\limits_0^1 {t^{\alpha  - 1} \left( {1 - t} \right)^{\beta  - 1} dt}$ are the Gamma and Beta functions, respectively.\normalsize
  \end{tablenotes}
 \end{threeparttable}	 	
\end{table*}

Practical ICs often contain also non-Gaussian parameters, and they cannot be effectively simulated by PC-based techniques. For such cases, gPC is more appealing since it can effectively handle non-Gaussian parameters. Motivated by this, Pulch applied gPC-based spectral methods to analyzing stochastic linear circuits~\cite{Pulch:2011}. Since almost all semiconductor devices are nonlinear, it is necessary to develop uncertainty quantification tools for nonlinear circuit simulation. Some progress has been reported along this line~\cite{zzhang:tcad2013,zzhang:tcas2_2013,Pulch:2011_1,Pulch:2009}. The RF circuit simulators in~\cite{Pulch:2011_1,Pulch:2009} directly apply gPC and stochastic Galerkin, showing remarkable speedup over MC. In order to further reduce the computational cost, the authors of this paper have proposed to simulate nonlinear circuits using a stochastic testing scheme~\cite{zzhang:tcad2013,zzhang:tcas2_2013}. Stochastic testing can be regarded as a hybrid version of stochastic collocation and stochastic Galerkin methods, and it proves more efficient for time-domain circuit simulation.

In this paper, we aim to review the fundamental ideas of gPC-based transistor-level simulation, and to summarize the recent progress on this topic. In Section II, we review some backgrounds on gPC and numerical quadrature. Section III discusses stochastic testing, stochastic Galerkin and stochastic collocation techniques and compares their performances in circuit simulation. In Section IV, some stochastic periodic steady-state simulators based on intrusive solvers are discussed and compared. Section V discusses some open problems in this field, followed by the conclusion in Section IV. 

\section{Preliminaries}
Consider the stochastic differential algebraic equation obtained from modified nodal analysis~\cite{mna:1975}:
\begin{equation}
\label{eq:sdae}
\begin{array}{l}
 \displaystyle{\frac{{d\vec q\left( {\vec x( {t,\vec \xi } ),\vec \xi } \right)}}{{dt}} }+ \vec f\left( {\vec x( {t,\vec \xi } ),\vec \xi } \right) = B\vec u\left( t \right) 
 \end{array}
\end{equation}
where $\vec u(t)$ is the input; ${\vec x}\in \mathbb{R}^n$ denotes nodal voltages and branch currents; ${\vec q}\in \mathbb{R}^n$ and ${\vec f}\in \mathbb{R}^n$ represent the charge/flux and current/voltage terms, respectively. Here ${\vec \xi}$=$[\xi_1,\cdots,\xi_d]\in\Omega$ (with $\Omega\subseteq\mathbb{R}^d$) denotes $d$ Gaussian and/or non-Gaussian parameters describing the device-level variations. Assume that all random parameters are independent, i.e., their  joint probability density function (PDF) can be expressed as
\begin{equation}
\label{PDF}
\rho(\vec \xi)=\prod\limits_{k = 1}^d {\rho _{k } \left( \xi_k \right)},
\end{equation}
with ${\rho _{k } \left( \xi_k \right)}$ being the PDF of $\xi_k \in \Omega_k$. In this paper, we focus on how to solve~(\ref{eq:sdae}) by stochastic spectral methods to extract the statistical information of the state vector $\vec x( {t,\vec \xi } )$. 

\subsection{Generalized Polynomial Chaos (gPC) Construction}
%We first summarize the procedures of constructing gPC basis functions.

\textbf{Univariate gPC.} For $\xi_k \in\Omega_k \subseteq \mathbb{R}$, one can construct a set of polynomial functions subject to the orthonormal condition:
\begin{equation}
\label{uni_gPC}
%\begin{array}{l}
 \left\langle {\phi^k_{\gamma} ( {\xi_k } ),\phi^k_{\nu} ( {\xi_k } )} \right\rangle  = \int\limits_{\Omega_k}  {\phi^k_{\gamma} ( {\xi_k } )\phi^k_{\nu} ( {\xi_k } ){\rho_k}( {\xi_k } )d\xi_k }=\delta_{\gamma,\nu} 
 %\end{array} \nonumber
\end{equation}
where $\langle , \rangle$ denotes the inner product; $\delta_{\gamma,\nu}$ is a Delta function; integers $\gamma$ and $\nu$ are the degrees of $\xi_k$ in polynomials $\phi^k_{\gamma} ( {\xi_k } )$ and $\phi^k_{\nu} ( {\xi_k } )$, respectively. Given ${\rho_k}( {\xi_k } )$ and $\Omega_k$, one can utilize a three-term recurrence relation to construct such orthonormal polynomials~\cite{Walter:1982}.
Some univariate gPC basis functions are listed in Table~\ref{tab:gPC} as a demonstration. It is worth noting that: 1) the univariate gPC basis functions are not limited to the cases listed in Table~\ref{tab:gPC}; 2) when $\xi_k$ is a Gaussian variable, its gPC simplifies to the Hermite polynomial chaos~\cite{PC1938}.

\textbf{Multivariate gPC.} When the components of $\vec \xi$ are assumed mutually independent, the multivariate gPC can be constructed based on the univariate gPC of each $\xi_k$. Given an index vector $\vec \alpha=[\alpha_1,\cdots,\alpha_d]\in \mathbb{N}^d $, the corresponding multivariate gPC is constructed as
\begin{equation}
H_{\vec \alpha} ( {\vec \xi } ) = \prod\limits_{k = 1}^d {\phi^k _{\alpha_k } ( {\xi _k } )}. 
\end{equation}
The obtained multivariate gPC is orthonormal, i.e.,
\begin{equation}
\label{multi_gPC}
 \left\langle {H_{\vec \alpha} ( {\vec \xi } ),H_{\vec \beta} ( {\vec \xi} )} \right\rangle  = \int\limits_{\Omega}  {H_{\vec \alpha} ( {\vec \xi } )H_{\vec \beta} ( {\vec \xi} ){\rho}( {\vec \xi } )d{\vec \xi} }=\delta_{\vec \alpha,\vec \beta}.
 \nonumber
\end{equation}
Note that $H_{\vec \alpha} ( {\vec \xi } ) $ is the product of different types of univariate gPC bases when $\xi_k$'s have different density functions.

\subsection{gPC Expansion}
If $\vec x({\vec \xi},t)$ is a 2nd-order stochastic process (i.e., $\vec x({\vec \xi},t)$ has a bounded 2nd-order moment), we can approximate it by a finite-term gPC expansion
\begin{equation}	
\label{gpcExpan}
\vec x(t,\vec \xi ) \approx \tilde x(t,\vec \xi) =\sum\limits_{\vec \alpha \in {\cal P}} {\hat x_{\vec \alpha} (t)H_{\vec \alpha}(\vec \xi )} 
\end{equation}
where $\hat x_{\vec \alpha} (t)\in \mathbb{R}^n$ denotes the gPC coefficient with index ${\vec \alpha}$, and ${\cal P}$ is a set containing some properly selected index vectors. 

Given $p\in \mathbb{N}^+$, there are two popular choices for ${\cal P}$~\cite{sgscCompare}. In the tensor product method, one sets ${\cal P}=\{\vec \alpha |\; 0\leq \alpha _k \leq p\}$, leading to a total of $(p+1)^d$ gPC bases. In order to reduce the total number of basis functions, the total degree scheme sets ${\cal P}=\{\vec \alpha |\; \alpha_k\in \mathbb{N},\; 0\leq {\alpha _1}+\cdots+\alpha_d \leq p\}$, leading to
\begin{equation}
\label{Kvalue}
K = \left( \begin{array}{l}
 p + d \\ 
 \;\;p \\ 
 \end{array} \right) = \frac{{(p + d)!}}{{p!d!}}
\end{equation}
gPC bases in total. This total degree method is employed in our stochastic circuit simulator. There is a one-to-one correspondence between $k$ (with $1\leq k\leq K$) and the index vector $\vec \alpha$, thus for simplicity (\ref{gpcExpan}) is normally rewritten as
\begin{equation}	
\label{gpcExpan_k}
\vec x(t,\vec \xi ) \approx \tilde x(t,\vec \xi) =\sum\limits_{k=1}^K {\hat x^k (t)H_k(\vec \xi )}.
\end{equation}

It is shown that gPC expansions converge exponentially for some analytical functions~\cite{book:Dxiu,gPC2002,xiu2009}. Such exponential convergence rates may not be observed in practical engineering problems, but gPC still converge very fast when the function of interest has a smooth dependence on $\vec \xi$. With gPC approximations, some statistical information (e.g., mean and variance) can be easily calculated due to the orthonormality of $H_{k}(\vec \xi)$'s.  
\begin{figure*}[t]
	\centering
		\includegraphics[width=150mm]{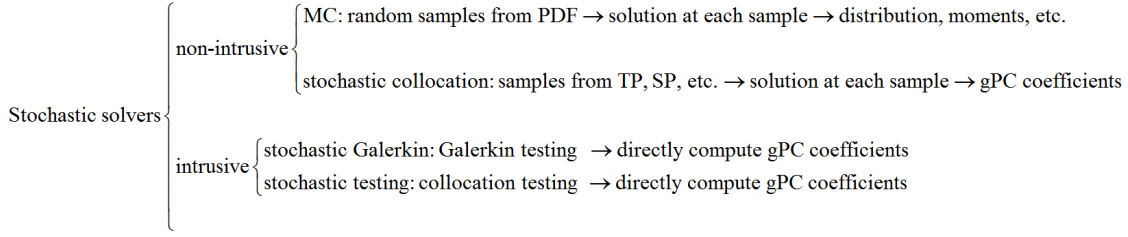} 
\caption{Classification of various stochastic solvers~\cite{zzhang:tcad2013}. ``TP" and ``SP" means the quadrature rules based on tensor product and sparse grids, respectively.}
	\label{fig:method_class}
\end{figure*}  

\subsection{Numerical Quadrature}
This section briefly reviews some numerical quadrature methods widely used in stochastic spectral methods.

\textbf{1-D Case.} When computing an integral with a quadrature method one typically uses the expression
\begin{equation}
\label{stoInt}
\int\limits_{\Omega _k } {g( {\xi _k } )\rho_k ( {\xi _k } )d\xi _k }  \approx \sum\limits_{j = 1}^{\hat n} {g( {\xi _k^j } )} w_k^j
\end{equation}
when $g\left( {\xi _k } \right)$ is a smooth function. The quadrature points ${\xi _k^j }$'s and weights $w_k^j$'s are chosen according to $\Omega_k$ and $\rho_k \left( {\xi _k } \right)$. Two kinds of quadrature rules are widely used: Gauss quadrature~\cite{Golub:1969} and Clenshaw-Curtis rules~\cite{Clenshaw:1960,Trefethen:2008}. With $\hat n$ points, Gauss quadrature rule produces exact results for all polynomials of degree $\leq 2\hat n-1$, and Clenshaw-Curtis gets exact results when the degree of $g(\xi_k)$ is $\leq \hat n-1$. Clenshaw-Curtis scheme generates nested quadrature points and assumes that $\xi_k$ is uniformly distributed in a bounded domain. 

\textbf{Multi-dimensional Case.} One can also evaluate a multidimensional integral in $\Omega\subseteq\mathbb{R}^d$ using the formula
\begin{equation}
\label{stoInt:md}
\int\limits_{\Omega } {g( {\vec \xi } ){\rho} ( {\vec \xi })d\vec \xi }  \approx \sum\limits_{j = 1}^{\hat N} {g( {\vec \xi ^j } )} w^j.
\end{equation}
where $\hat N$ is the total number of quadrature points, and $w^j$ is the weight corresponding to quadrature point $\vec \xi ^j$. Given the 1-D quadrature points for each $\xi _k$, $\vec \xi^j$'s and $w^j$'s can be obtained for instance using a tensor-product rule or using sparse grids~\cite{sparse_grid:2000,Gerstner:1998}. With Smolyak's algorithm, sparse grid technique uses much fewer quadrature points than the tensor-product rule, thus it is widely used to solve stochastic PDEs~\cite{col:2005,Ivo:2007,Nobile:2008,Nobile:2008_2}. In~\cite{col:2005,Ivo:2007,Nobile:2008,Nobile:2008_2} Smolyak's algorithm produces nested sparse grids because all random parameters are assumed uniformly distributed (and thus Clenshaw-Curtis rule is used for all $\xi_k$'s). However, Smolyak's algorithm generates non-nested sparse grids when non-nested 1-D quadrature points are used for some parameters (since many random parameters with non-uniform distributions may not be effectively handled by the Clenshaw-Curtis rule).

\section{Stochastic Spectral Methods}

\subsection{Classification of Stochastic Solvers}
The main stochastic solvers are classified in Fig.~\ref{fig:method_class}. MC and stochastic collocation are both non-intrusive (or sampling-based) methods: they solve (\ref{eq:sdae}) as a deterministic problem at a set of samples. Their main difference lies in the sampling stage: MC randomly draws some samples based on $\rho(\vec \xi)$, whereas stochastic collocation typically uses the points from a tensor-product or sparse-grid rule such that the gPC coefficients can be well reconstructed. Stochastic Galerkin and stochastic testing belong to the family of intrusive solvers: through solving a new coupled differential algebraic equation they directly compute the gPC coefficients. The former sets up the coupled equation by Galerkin testing, whereas the latter constructs a coupled equation via collocation testing. %Such a collocation testing formulation leads to significant speedup.

\subsection{Stochastic Testing (ST)}
\label{subsec:st}
The stochastic testing method needs to select $K$ testing points $\vec \xi_1, \cdots, \vec \xi_K$. First, a quadrature scheme (e.g., tensor-product or sparse grid rule in Section II-C) is applied to generate $\hat N$ quadrature points $\vec \xi^j$'s in parameter space $\Omega$, which are called candidate nodes. Second, the $K$ most important candidate nodes are selected such that the transformation matrix $\textbf{V}\in \mathbb{R}^{K\times K}$, with its $(i,j)$ entry being 
\begin{equation}
\label{def:V}
\textbf{V}_{i,j} = {H_j (\vec \xi _i )},
\end{equation}
is invertible and well conditioned. %This is completed by checking the rank of $\textbf{V}$ on the fly~\cite{zzhang:tcad2013}.
\begin{algorithm}[t]
\caption{Testing Point Selection for ST~\cite{zzhang:tcad2013}.}
\label{alg:testNode}
\begin{algorithmic}[1]
\STATE {construct $\hat N$ $d$-D quadrature points and weights; }
\STATE {reorder the quadrature points such that $|w^j|\geq |w^{j+1}|$;} 
\STATE {set $V=\vec H\left( \vec \xi^1\right)/||\vec H\left( \vec \xi^1\right) ||$, $\vec \xi_1=\vec \xi^1$, and $m=1$; } 
%\STATE {set $\vec \xi_1=\vec \xi^1$, and $m=1$; } %\;\; \%      the 1st testing node} 
\STATE {\textbf{for} $j=2,\;\cdots$, $\hat N$ \textbf{do}}
 \STATE {\hspace{10pt} $\vec v=\vec H\left( \vec \xi^j\right)-V\left( V^T\vec H\left( \vec \xi^j\right)\right)$;}
 \STATE {\hspace{10pt}\textbf{if} $||\vec v||/||\vec H\left( \vec \xi^j\right) ||>\beta$}
   \STATE {\hspace{20pt}set $V=[V;\vec v/||\vec v||]$, $m=m+1$, $\vec \xi_m=\vec \xi^j$;}
   %\STATE {\hspace{20pt} $\vec \xi_m=\vec \xi^j$; \;\; \% select as a new testing node.}
    \STATE {\hspace{20pt}\textbf{if} $m\geq K$, break, \textbf{end};}
     \STATE {\hspace{10pt}\textbf{end if}}
     \STATE {\textbf{end for} } 
\end{algorithmic}
\end{algorithm}
Define a vector function $\vec H(\vec \xi):=[H_1(\vec \xi); \cdots; H_K(\vec \xi)]$, then the testing points can be selected by Algorithm~\ref{alg:testNode}~\cite{zzhang:tcad2013}. Only a small portion of the candidate nodes are finally selected as the testing points. 

Let $\hat{\textbf{x}}(t)=[\hat x^1(t);\cdots ;\hat x^K(t)]$ denote the gPC coefficients, $\tilde q(\hat{\textbf{x}}(t),\vec \xi)=\vec q( {\tilde x( {t,\vec \xi } ),\vec \xi } )$ and $\tilde f(\hat{\textbf{x}}(t),\vec \xi)=\vec f( {\tilde x( {t,\vec \xi } ),\vec \xi } )$. Substituting $\tilde x(t,\vec \xi)$ of (\ref{gpcExpan_k}) into (\ref{eq:sdae}) yields a residual function
\begin{equation}
\label{eq:residual}
\begin{array}{l}
 {\rm R}(\hat{\textbf{x}}(t),\vec \xi )=\displaystyle{\frac{{d\tilde q(\hat{\textbf{x}}(t),\vec \xi)}}{{dt}}} + \tilde f(\hat{\textbf{x}}(t),\vec \xi)- B\vec u( t ).
 \end{array}
\end{equation}

\textbf{Collocation Testing.} Enforcing the residual function to zero at all testing points, stochastic testing generates the following coupled differential algebraic equation:
\begin{align}
\label{ST:forced}
\frac{{d\textbf{q}(\hat{\textbf{x}}(t))}}{{dt}} + \textbf{f}(\hat{\textbf{x}}(t)) = \textbf{B}u(t),
\end{align}
where the $k$-th blocks of $\textbf{q}(\hat{\textbf{x}}(t))$, $\textbf{f}(\hat{\textbf{x}}(t))$ and $\textbf{B}$ are $\tilde q(\hat{\textbf{x}}(t),\vec \xi_k)$, $\tilde f(\hat{\textbf{x}}(t),\vec \xi_k)$ and $B$, respectively. 

\textbf{Numerical Solver.} Stochastic testing is an intrusive solver: the gPC coefficients $\hat{\textbf{x}}(t)$ are directly computed by simulating (\ref{ST:forced}), then the parameter-dependent current/voltage variables are obtained by gPC approximations. In transient analysis, the time step sizes can be selected adaptively according to the local truncation error (LTE) of (\ref{ST:forced}) as done in commercial deterministic circuit simulators~\cite{kundertbook:1995,Tuinenga:1995}. Another desirable feature of stochastic testing is the decoupling procedure inside the intrusive solver. Assume that $\textbf{J}$ is the Jacobian inside the Newton's iteration when simulating (\ref{ST:forced}) (as a DC problem or as a transient problem using numerical integration such as backward Euler). It is shown in~\cite{zzhang:tcad2013} that $\textbf{J}$ can be factored as
\begin{equation}
\textbf{J}={\rm blkdiag}(J_1,\cdots, J_K)(\textbf{V}\otimes\textbf{I}_n)
\end{equation}
where ${\rm blkdiag}$ is the block diagonal operator, $\otimes$ is the Kronecker product operation, and $\textbf{I}_n \in \mathbb{R}^{n\times n}$ is an identity matrix. Matrix $J_k\in \mathbb{R}^{n\times n}$ can be treated as a Jacobian corresponding to (\ref{eq:sdae}) with $\vec \xi=\vec \xi_k$. Since the Vandermonde-like matrix \textbf{V} [as defined in (\ref{def:V})] can be easily inverted~\cite{fastInverse}, the linear system solution inside each Newton's iteration can be decoupled into $K$ small-size problems. Consequently, the overall computational cost scales linearly with $K$~\cite{zzhang:tcad2013}. %Such a procedure leads to significant speedup and allows potentially parallel computation.

\subsection{Stochastic Galerkin (SG)}
\label{subsec:sg} 
\textbf{Galerkin Testing.} Applying Galerkin testing
\begin{equation}
\left <{\rm R}(\hat{\textbf{x}}(t),\vec \xi ), H_k(\vec \xi)\right >=\int\limits_{\Omega } {{\rm R}(\hat{\textbf{x}}(t),\vec \xi ) H_k(\vec \xi){\rho} ( {\vec \xi })d\vec \xi }=0
\end{equation}
for $k=1,\cdots, K$, stochastic Galerkin forms a coupled equation in the form of (\ref{ST:forced}). Now the $k$-th blocks of $\textbf{q}(\hat{\textbf{x}}(t))$, $\textbf{f}(\hat{\textbf{x}}(t))$ and $\textbf{B}$ are $\left<\tilde q(\hat{\textbf{x}}(t),\vec \xi), H_k(\vec \xi)\right >$, $\left <\tilde f(\hat{\textbf{x}}(t),\vec \xi),H_k(\vec \xi) \right >$ and $\left < B, H_k(\vec \xi) \right >$, respectively. The inner products can be evaluated using the numerical quadrature rules described in Section II-C or by an MC integration (if $d$ is large).

\textbf{Numerical Solver.} After (\ref{ST:forced}) is formed by Galerkin testing, $\hat{\textbf{x}}(t)$ is also computed in an intrusive manner. In time-domain simulation, the time step sizes can also be controlled adaptively as in stochastic testing. Compared with stochastic testing, stochastic Galerkin has two drawbacks. First, the inner product evaluation needs $\hat N>K$ quadrature points, and thus at each time point stochastic Galerkin requires more circuit/device evaluations. This can lead to remarkable time cost when complex semiconductor device models are employed. Second, the resulting Jacobian in a stochastic Galerkin-based simulator cannot be decoupled, although it can be decoupled for linear circuits if the gPC bases are chosen by the tensor product method~\cite{Pulch:2011}. This causes a significant computational overhead compared with stochastic testing.

\subsection{Stochastic Collocation (SC)}
In stochastic collocation, Eq. (\ref{eq:sdae}) is first is solved at $\hat N$ sample points to obtain a set of deterministic solutions $\vec x(t,\vec \xi^k)$'s. After that the gPC coefficients are reconstructed by a post-processing numerical scheme. In the mainstream stochastic collocation schemes~\cite{col:2005,Ivo:2007,Nobile:2008,Nobile:2008_2}, the samples are selected by a tensor product or sparse-grid quadrature technique, and thus the $j$-th gPC coefficient vector can be estimated by
\begin{equation}
\label{SC:interpolation}
\hat x^{j } ( t ) = \left\langle {\vec x ( t,{\vec \xi } ),H_{j } ( {\vec \xi } )} \right\rangle\approx \sum\limits_{k = 1}^{\hat N } {w^k H_{j } ( {\vec \xi ^k } )} \vec x(t, {\vec \xi ^k } ).
\end{equation}
%Other technique such as Lagrange's interpolation can also be utilized to extract the gPC coefficients.

In practical time-domain simulation, each $x(t,\vec \xi^k)$ is computed at a set of discretized time points. Therefore, to reconstruct the gPC coefficients, the deterministic solutions for all samples should be located on the same time grid. Since it is difficult to preselect an adaptive time grid for the black-box deterministic solver, a small fixed step size is normally used, leading to excessive computational cost for stiff circuits. 

The speedup factor of stochastic testing over stochastic collocation can be estimated as~\cite{zzhang:tcad2013}
\begin{equation}
\kappa_{\rm overall}=\kappa_{\rm samp}\times \kappa_{\rm tctrl}.
\end{equation}
Here $\kappa_{\rm samp}={\hat N}/{K}>1$ because stochastic testing uses fewer samples than stochastic collocation. If stochastic collocation uses tensor-product quadrature points, $\kappa_{\rm samp}$ gets extremely large as $d$ increases. When stochastic collocation uses nested Smolyak sparse grids and the total degree of the gPC expansion is $p$, $\kappa_{\rm samp}$ is about $2^p$ for $d\gg 1$. The second factor $\kappa_{\rm tctrl}>1$ is caused by adaptive time stepping in stochastic testing, which is case dependent. In DC analysis, $\kappa_{\rm tctrl}=1$.

\subsection{Performance Analysis}
%ST, SG and SC are implemented in MATLAB, and different simulations (DC, transient, AC) are performed on several analog/RF and digital ICs with both Gaussian 
We have implemented stochastic testing, stochastic Galerkin and stochastic collocation in MATLAB and performed various simulations (DC, transient, AC) on several analog/RF and digital ICs~\cite{zzhang:tcad2013}. For those benchmarks with several Gaussian and non-Gaussian random parameters, all stochastic spectral methods have shown $10^2$--$10^3\times$ speedup over MC due to the fast convergence of gPC expansions. The speedup factors of stochastic testing over stochastic Galerkin and stochastic collocation are on the level of $O(1)$ to $O(10^2)$, which are more significant as the gPC order $p$ increases.
\begin{figure}[t]
	\centering
		\includegraphics[width=3.3in]{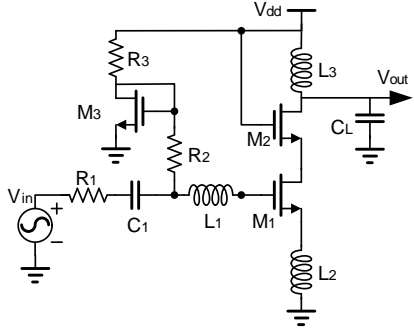} 
\caption{Schematic of the LNA.}
	\label{fig:LNA}
\end{figure} 
\begin{figure}[t]
	\centering
		\includegraphics[width=3.3in]{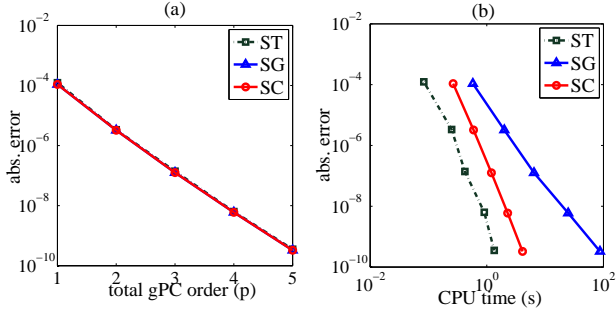} 
\caption{Accuracy and efficiency of stochastic testing (ST), stochastic Galerkin (SG) and stochastic collocation (SC) for the DC analysis of LNA.} %: (a) $L_2$ errors versus $p$, (b) $L_2$ errors versus CPU times.
	\label{fig:methodCompare_LNA}
\end{figure}

The static analysis of a common-source amplifier (with four random parameters) shows that stochastic testing has slightly larger errors than stochastic Galerkin and stochastic collocation, but it uses the least amount of CPU time to achieve a similar level of accuracy. The results of a low-noise amplifier (LNA) with three random parameters (in Fig.~\ref{fig:LNA}) is plotted in Fig.~\ref{fig:methodCompare_LNA}. The $L_2$-norm errors of the computed gPC coefficients from all three methods are almost the same, and stochastic testing costs significantly less CPU time. Our experiments show that a $3$rd-order gPC expansion (i.e., $p=3$) is enough for most circuits.
\begin{figure}[t]
	\centering
		\includegraphics[width=3.3in]{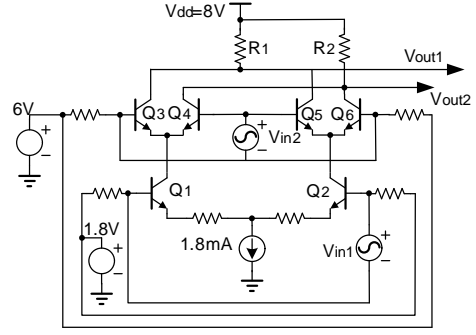} 
\caption{Schematic of the BJT double-balanced mixer.}
	\label{fig:dbmixer}
\end{figure} 
\begin{figure}[t]
	\centering
		\includegraphics[width=85mm]{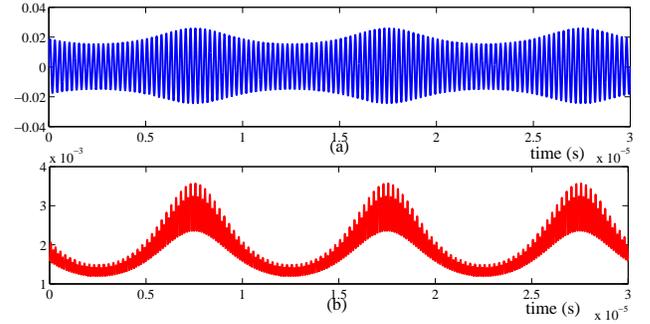} 
\caption{Uncertainties of $V_{\rm out}$$=$$V_{{\rm out}1}-V_{{\rm out}2}$ of the double-balanced mixer: (a) mean value, (b) standard deviation.}
	\label{fig:dbmixer_output}
\end{figure}  

Transient simulation of the common-source amplifier and LNA shows that the speedup factor of stochastic testing over stochastic Galerkin and stochastic collocation is about $O(10^1)$ to $O(10^2)$, which is more significant for large-size circuits. In analog circuits, the speedup factor caused by adaptive time stepping is about $O(1)$ to $O(10^1)$. For digital (e.g., SRAM cell) and multi-rate RF (e.g., BJT mixer) circuits, stochastic testing can solve the problem with seconds or minutes of CPU time, whereas stochastic collocation may require $>1$ hour due to the uniform time stepping. Fig.~\ref{fig:dbmixer} shows a mixer with uncertainties at $R_1$ and $R_2$. Stochastic testing produces the mean and standard-deviation waveforms (in Fig.~\ref{fig:dbmixer_output}) after $21$ minutes, whereas stochastic Galerkin, stochastic collocation and MC are prohibitively expensive on the MATLAB platform.

\section{Uncertainty Quantification for Periodic Steady States}
Analog/RF and power electronic circuit designers are interested in periodic steady-state analysis~\cite{kundert:jssc99,Nastov:ieeeProc,Jacob:matrixfree,Aprille:ieeeProc,Aprille:TCAS,Vytyaz:tcad}. Using stochastic spectral methods, uncertainties of the periodic steady states can be analyzed more efficiently than using MC. This section summarizes the progress on stochastic time-domain periodic steady-state solvers~\cite{zzhang:tcas2_2013}. Other solvers (e.g., harmonic balance) can also be easily implemented.

\subsection{Forced Circuits}
For many forced circuits (e.g., amplifiers and power converters), there exists a periodic steady-state solution $\vec x(t,\vec \xi)=\vec x(t+T, \vec \xi)$ when the input is a time-varying periodic signal $\vec u(t)=\vec u(t+T)$. The state vector $\vec x(t,\vec \xi)$ is periodic for any $\vec \xi\in \Omega$ if and only if $\hat{\textbf{x}}(t)$ is periodic. Therefore, we can set up the following equation
\begin{align}
\label{st_pss_forced}
 \textbf{g} (\hat{\textbf{y}}) =  \Phi (\hat{\textbf{y}},0,T) - \hat{\textbf{y}} = 0.
\end{align}
Here $\hat{\textbf{y}}=\hat{\textbf{x}}(0)$, and $\hat{\textbf{x}}(T)=\Phi(\hat{\textbf{y}},0,T)$ is the state transition function of (\ref{ST:forced}) formed by stochastic testing (c.f. Section~\ref{subsec:st}) or stochastic Galerkin (c.f. Section~\ref{subsec:sg}). 

Eq. (\ref{st_pss_forced}) can be solved by the standard shooting Newton method~\cite{kundert:jssc99,Nastov:ieeeProc,Jacob:matrixfree,Aprille:ieeeProc}. When solving the linear equation inside each Newton's iteration, evaluating the right-hand side requires integrating (\ref{ST:forced}) from $t=0$ to $t=T$, and the Jacobian matrix can be obtained once a Monodromy matrix is computed (via a sensitivity analysis along the discretized trajectories). Directly solving (\ref{st_pss_forced}) requires $O(K^3n^3)$ cost if a direct matrix solver is employed. Fortunately, \cite{zzhang:tcas2_2013} shows that the linear equation solution can be easily decoupled into $K$ small problems by a similarity transform, if (\ref{ST:forced}) is formed by stochastic testing. The decoupled intrusive transient solver in Section~\ref{subsec:st} can be employed to evaluate the right-hand side of each linear equation inside Newton's iterations, thus the overall cost can be reduced to $KO(n^3)$ in the stochastic testing formulation.

\begin{figure}[t]
	\centering
		\includegraphics[width=3.3in]{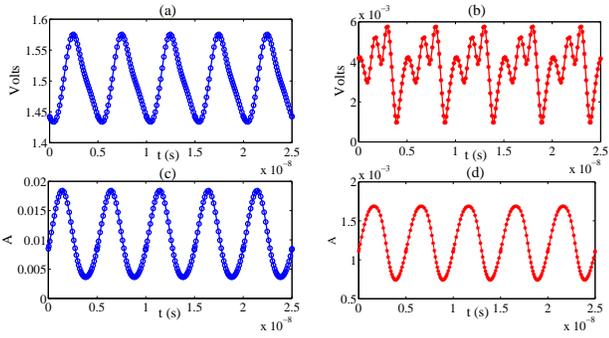} 
\caption{Periodic steady-state waveforms for the LNA. (a) $\&$ (b): mean and s.t.d of $V_{\rm out}$; (c) $\&$ (d): mean and s.t.d of $I (V_{\rm dd})$.}
	\label{fig:LNA_wave}
\end{figure}

\begin{figure}[t]
	\centering
		\includegraphics[width=3.3in]{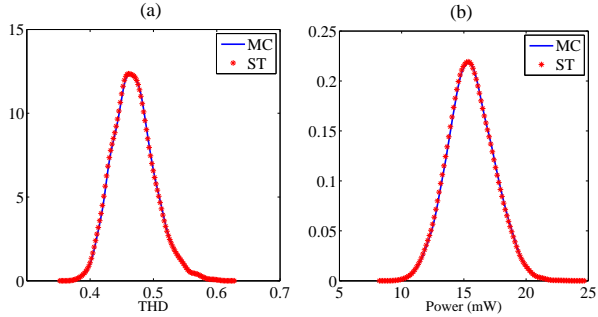}
\caption{Probability density functions obtained by MC and stochastic testing (ST). (a) total harmonic distortion and (b) power dissipation.}
	\label{fig:LNA_pdf}
\end{figure}

\textbf{Results.} The simulation result of the LNA (with $V_{\rm in}=0.1{\rm sin}(4\pi\times 10^8t)$ V) is plotted in Fig.~\ref{fig:LNA_wave}. With a $3$rd-order total-degree gPC expansion, the stochastic testing-based and stochastic Galerkin-based solvers give the same results. Using a standard MC, $8000$ samples are required to achieve a similar level of accuracy ($<$$1\%$ relative errors for the mean and standard deviation). Fig.~\ref{fig:LNA_pdf} plots the density functions of the total harmonic distortion and power consumption extracted from the computed periodic steady-state solution, which are consistent with those from MC. The simulation cost of the decoupled stochastic testing solver is $3.4$ seconds, which is $42\times$ faster over the coupled stochastic testing solver, $71\times$ faster over the stochastic Galerkin-based solver, and $220\times$ faster over MC. In~\cite{zzhang:tcas2_2013} an $O(K^2)$ speedup factor caused by decoupling is clearly observed for stochastic testing.

\subsection{Autonomous Circuits}
For unforced cases (e.g., oscillators), the input signal $\vec u(t)=\vec u$ is time-invariant, and the period is unknown. The periodicity constraint is $\hat{\textbf{x}}(t,\vec \xi)=\hat{\textbf{x}}(t+T(\vec \xi),\vec \xi)$, where the period $T(\vec \xi)$ depends on $\vec \xi$. Choose a constant $T_0$ and assume that $a(\vec \xi)$ is a scaling factor such that $T(\vec \xi)=T_0 a(\vec \xi )$, then we obtain a scaled time variable $\tau=t/a (\vec \xi )$~\cite{Vytyaz:tcad}. Let $\vec z(\tau,\vec{\xi}):=\vec x (t,\vec \xi)$, then $\vec z(\tau,\vec \xi)$ has a constant period $T_0$ on the scaled time axis $\tau$. Both $a(\vec \xi)$ and $\vec z(\tau, \vec \xi)$ can be approximated by gPC expansions
\begin{equation}
\begin{array}{l}
 a(\vec \xi ) \approx \tilde a(\vec \xi ) = \sum\limits_{k = 1}^K {\hat a^k H_k (\vec \xi )} ,\;\; \\ 
 \vec z(\tau ,\vec \xi ) \approx \tilde z(\tau ,\vec \xi ) = \sum\limits_{k = 1}^K {\hat z^k (\tau )H_k (\vec \xi )} .  
 \end{array}
\end{equation}

Substituting the above approximation into (\ref{eq:sdae}) and changing the time variable, we obtain a new residual function
\begin{equation}
\label{eq:residual_tau}
\begin{array}{l}
 {\rm R}(\hat{\textbf{z}}(\tau),\hat {\textbf{a}},\vec \xi )=\displaystyle{\frac{{d\tilde q(\hat{\textbf{z}}(\tau),\vec \xi)}}{{d\tau}}} + \tilde a(\vec \xi)\tilde f(\hat{\textbf{z}}(\tau),\vec \xi)- \tilde a(\vec \xi)B\vec u.
 \end{array} \nonumber
\end{equation}
Here $\tilde q(\hat{\textbf{z}}(\tau),\vec \xi)=\vec q(\tilde{z}(\tau,\vec \xi),\vec \xi)$, $\tilde f(\hat{\textbf{z}}(\tau),\vec \xi)=\vec f(\tilde{z}(\tau,\vec \xi),\vec \xi)$; $\hat {\textbf{a}}$ and $\hat{\textbf{z}}(\tau)$ collect the gPC coefficients of $\tilde a(\vec \xi )$ and  $\tilde z(\tau ,\vec \xi )$, respectively. The following coupled differential equation
\begin{equation}
\label{ST:unforced}
\displaystyle{\frac{{d\textbf{q}\left(\hat{\textbf{z}}(\tau)\right)}}{{d\tau}}} + \textbf{f}\left(\hat{\textbf{z}}(\tau),\hat{\textbf{a}}\right)= \textbf{B}(\hat{\textbf{a}})\vec u
 %\end{array}
\end{equation}
can be constructed by either stochastic testing~\cite{zzhang:tcas2_2013} or stochastic Galerkin~\cite{Pulch:2011_1}. In stochastic testing we perform collocation testing (c.f. Section~\ref{subsec:st}) on  ${\rm R}(\hat{\textbf{z}}(\tau),\hat {\textbf{a}},\vec \xi )$, whereas in stochastic Galerkin one applies Galerkin testing (c.f. Section~\ref{subsec:sg}). %In ST: in ST, $ {\rm R}(\hat{\textbf{z}}(\tau),\hat {\textbf{a}},\vec \xi )$ is enforced to zero at all testing points~\cite{zzhang:tcas2_2013}, whereas SG enforces ${\rm R}(\hat{\textbf{z}}(\tau),\hat {\textbf{a}},\vec \xi )$ orthogonal to all $K$ basis functions used in the gPC expansion~\cite{Pulch:2011_1}. In ST, the $k$-th blocks of $\textbf{q}\left(\hat{\textbf{z}}(\tau)\right)$, $\textbf{f}\left(\hat{\textbf{z}}(\tau),\hat{\textbf{a}}\right)$ and $\textbf{B}(\hat{\textbf{a}})$ are $\tilde q\left(\hat{\textbf{z}}(\tau),\vec \xi_k\right)$, $\tilde a(\vec \xi_k)\tilde f\left(\hat{\textbf{z}}(\tau),\vec \xi_k\right)$ and $\tilde a(\vec \xi_k)B$, respectively, whereas in SG formulation they are evaluated by inner products as done in Section~\ref{subsec:sg}.

Based on (\ref{ST:unforced}), an algebraic equation can be set up to solve for the gPC coefficients of $\tilde z(0, \vec \xi)$ and $\tilde a(\vec \xi)$. Let $\hat{\textbf{y}}:=[\hat{\textbf{z}}(0);\hat {\textbf{a}}]$ and fix the $j$-th component of $\vec z(0)$ at $\lambda$, then we have
\small
\begin{equation}
\label{st_pss_unforced}
\textbf{g} ( {\hat{\textbf{y}} } ) = \left[ {\begin{array}{*{20}c}
   {\Psi ( {\hat{\textbf{z}}(0) ,\hat{\textbf{a}}} )}  \\
   {\chi ( \hat{\textbf{z}}(0))}  \\
\end{array}} \right] = \left[ {\begin{array}{*{20}c}
   {\Phi ( {\hat{\textbf{z}}(0),0,T_0 ,\hat{\textbf{a}}} ) - \hat{\textbf{z}}(0)}  \\
   {\chi ( \hat{\textbf{z}}(0))}  \\
\end{array}} \right] = 0.
\end{equation} \normalsize
Here $\Phi ( {\hat{\textbf{z}}(0),0,T_0 ,\hat{\textbf{a}}} )$ is the state transition function of (\ref{ST:unforced}), which depends on $\hat{\textbf{a}}$. The phase constraint $\chi ( \hat{\textbf{z}}(0))=0\in \mathbb{R}^K$ 
\begin{align}
\chi ( \hat{\textbf{z}}(0)) = \left[ \hat{\textbf{z}}_j(0) - \lambda ;\;  \hat{\textbf{z}}_{j + n}(0);\;  { \cdots ;\;}  \hat{\textbf{z}}_{j + (K - 1)n}(0)  \right] = 0 \nonumber
\end{align}
is added to make (\ref{st_pss_unforced}) a determined equation.

When solving (\ref{st_pss_unforced}) by Newton's iterations, the Jacobian evaluation is more involved than that in forced circuits. Besides the Monodromy matrix, the sensitivity matrix of $\textbf{g} ( {\hat{\textbf{y}} } )$ w.r.t $\hat{\textbf{a}}$ is also required, which can be obtained in an iterative way~\cite{zzhang:tcas2_2013}. Similar to the forced circuits, decoupling leads to an $O(K^2)$ speedup if the stochastic testing formulation is employed~\cite{zzhang:tcas2_2013}.
\begin{figure}[t]
	\centering
		\includegraphics[width=3.3in]{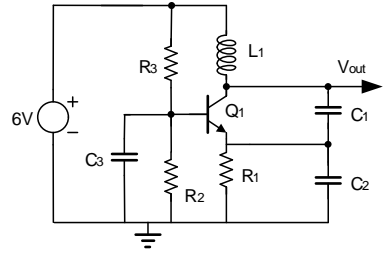} 
\caption{Schematic of the BJT Colpitts oscillator.}
	\label{fig:Colp_osc}
\end{figure} 
 \begin{figure}[t]
	\centering
		\includegraphics[width=3.3in]{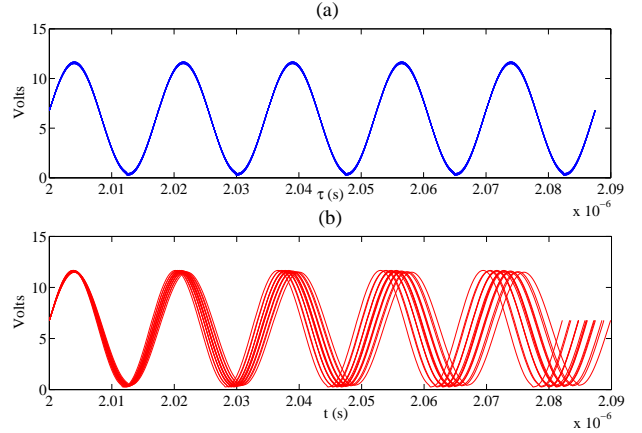} 
\caption{Realizations of $V_{\rm out}$ for the Colpitts oscillator. (a) on the scaled time axis, (b) on the original time axis.}
	\label{fig:Colp_wave}
\end{figure}

\textbf{Results.} The gPC-based periodic steady-state solvers are applied to analyze the BJT Colpitts oscillator in Fig.~\ref{fig:Colp_osc}. The oscillation frequency is influenced by the Gaussian variation of $L_1$ and non-Gaussian variation of $C_1$. With a $3$rd-order gPC expansion, the stochastic testing-based~\cite{zzhang:tcas2_2013} and stochastic Galerkin-based~\cite{Pulch:2011_1} solvers produce the same results. Fig.~\ref{fig:Colp_wave} shows some realizations of $V_{\rm out}$. The variation looks small on the scaled time axis $\tau$, but it is significant on the original time axis due to the uncertainties of the frequency. The CPU time of the decoupled stochastic testing-based solver is $4.9$ seconds, which is $2\times$ and $5\times$ faster over the coupled stochastic testing-based solver and the stochastic Galerkin-based solver~\cite{Pulch:2011_1}, respectively. To achieve the similar level of accuracy ($<1\%$ errors for the mean and standard deviation of the frequency), MC must use $5000$ samples, which is about $507\times$ slower than the stochastic testing-based simulator with decoupling.

\subsection{Other Related Work}
An intrusive simulator has been proposed to analyze the uncertainties of RF circuits with multi-rate input signals~\cite{Pulch:2009}. It uses the multi-time PDE technique~\cite{Jaijeet:2001} to solve a coupled differential equation formed by stochastic Galerkin, generating stochastic quasi-periodic steady-state solutions. The stochastic testing-based formulation can be easily extended to this case to further reduce the computational cost. 

Non-intrusive periodic steady-state solvers are not discussed in this paper due to their ease of implementation.

\section{Open Problems}
%Although stochastic spectral methods seem promising for stochastic circuit simulation, there are some important open problems as summarized (but not limited to) below. 
Although stochastic spectral methods seem promising for stochastic circuit simulation, there still exist many open problems, some of which are summarized below. 

\textbf{High Dimensionality.} The number of total gPC bases increases very fast as the parameter dimensionality $d$ increases. Consequently, the computational cost becomes prohibitively expensive when $d$ is large. It is worth exploiting the sparsity of the gPC coefficients to reduce the complexity. Compressed sensing~\cite{Donoho:2006} seems effective for behavior modeling~\cite{xli2010}, but its efficiency can degrade for simulation problems (since the gPC coefficients of different nodal voltages and/or branch currents have different sparsity pattens). A dominant singular vector method has been proposed for high-dimensional linear stochastic problems~\cite{Tarek_DAC:08}, yet solving the non-convex optimization is challenging for nonlinear problems.

\textbf{Correlated Non-Gaussian Parameters.} In existing literatures, the parameters are typically assumed mutually independent, which is not valid for many practical circuits. Unlike Gaussian variables, correlated non-Gaussian parameters cannot be easily transformed to independent ones, making the gPC basis construction challenging. A theoretical method has been proposed to deal with parameters with arbitrary density functions~\cite{arb_chaos}, but its numerical implementation is non-trivial.

\textbf{Long-Term Integration.} In digital IC simulation, normally designers have to perform a long-time transient simulation. In the applied math community, it is well known that PC/gPC approximation can be inaccurate for a tong-time integration, despite some improvements~\cite{Wan:2006}.

\section{Conclusion}
Stochastic spectral methods have emerged as a promising technique for the uncertainty quantification of integrated circuits. After reviewing some key concepts about gPC, this paper has discussed stochastic testing, stochastic Galerkin  and stochastic collocation methods, as well as their implementation and performance in nonlinear transistor circuit analysis. Some recent progress on stochastic periodic steady-state analysis has been summarized. Among these techniques, stochastic testing has shown higher efficiency in time-domain IC simulation. Some important problems, such as how to deal with high parameter dimensionality, correlated non-Gaussian parameters and long-term integration errors, have not been solved.


\section*{Acknowledgment}
This work was supported by the MI-MIT Collaborative Program (Reference No.196F/002/707/102f/70/9374). I. Elfadel's work was also supported by SRC under the MEES I, MEES II, and ACE$^{4}$S programs, and by ATIC under the TwinLab program. Z. Zhang would like to thank Dr. Tarek El-Moselhy for his helpful discussions during the work of~\cite{zzhang:tcad2013,zzhang:tcas2_2013}.



%\bibliographystyle{IEEEtran}
\bibliographystyle{IEEEtran}
\bibliography{date}

\end{document}